%Paper: gr-qc/9401014
%From: STEP8030@bureau.ucc.ie
%Date: 17 Jan 1994 10:39:25 +0000 (GMT)

%format plaine
\magnification=\magstep1

\centerline{\bf  Trapped surfaces and the Penrose inequality in spherically
symmetric geometries.}

\vskip 2cm
\centerline{ Edward Malec$^*$}
\centerline{ Niall \'O Murchadha}
\centerline{ Physics Department}
\centerline{University College}
\centerline{Cork, Ireland}
\vskip 2cm
We demonstrate that the Penrose inequality is valid for spherically
symmetric  geometries even when the
horizon is immersed in  matter. The matter field need not be at rest. The only
restriction is that the source   satisfies
the weak energy condition outside the horizon. No restrictions are placed on
the matter inside the horizon. The proof of the Penrose inequality
gives a new necessary  condition for the formation of trapped
surfaces. This formulation  can also be adapted to give a sufficient condition.
We show that a   modification  of the Penrose inequality proposed by Gibbons
for
charged black holes can be broken in early stages  of  gravitational collapse.
This investigation is based  exclusively on the initial  data formulation of
General Relativity.

\vskip 2cm
\centerline{$^*$(On leave of absence from the  Institute  of Physics,
Jagellonian University, 30-064 Krak\'ow, Poland)}

\vfill \eject
In our early analyses $^1$ of the formation of trapped surfaces in spherically
symmetric selfgravitating systems we found several  criteria that determine
the formation of trapped surfaces. These were all expressed in terms of
quasilocal  quantities. We found both necessary and sufficient conditions when
the matter was at rest (moment of time symmetry data) and a sufficient
condition when the matter was moving. We failed, however, to discover a
necessary condition in the case of non-symmetric in time initial data. In this
paper we fill that gap. We do this by first proving the Penrose inequality  in
the situation when the apparent horizon is inside  matter. The resulting
equation can be manipulated to give
 both   necessary and sufficient conditions
for the formation of trapped surfaces. These inequalities are interesting in
that they use a global quantity, the ADM mass.   The only assumptions we make
are that  a
 1+3 splitting of the spacetime by maximal  hypersurfaces exists and that the
matter density is nonnegative {\it outside} any horizon. Only a part of the
Einstein equations,  the hamiltonian and momentum constraints,   are employed.

The apparent horizon in the Schwarzschild geometry is a surface which
satisfies 2$m$/$R$ = 1, where $m$ is the total (ADM)  mass of the spacetime and
$R$ is the Schwarzschild (areal) radius of the surface. If the matter in a
spherically symmetric spacetime has compact support then the
metric in the exterior region can be written in the Schwarzschild form and
the condition 2$m$/$R$ = 1 is both a necessary and sufficient condition for the
appearance of trapped surfaces even in the case where the matter is
moving.   In this note we derive inequalities which are direct
generalizations of the 2$m$/$R$ = 1 condition for the appearance of a horizon
which are valid even when the surface in question is inside the support of the
matter as a follow-on from a proof of the Penrose inequality in such
circumstances.

 Penrose proposed$^2$ an inequality which he hoped would be satisfied by black
holes. This inequality reduces to the condition 2$m$/$R$ = 1 for the horizon
of a Schwarzschild black hole. The Penrose inequality  and its generalization
describing charged matter  proposed  by Gibbons $^3$    was formulated in order
to clarify circumstances in  which  the cosmic censorship hypothesis $^4$ can
be
broken.  Ludvigsen and Vickers $^5$ have proven that the  Penrose inequality
holds for a class of (possibly nonspherical)  geometries, assuming a global
condition on the past history of a collapsing system.  There are also several
partial proofs $^{6,7}$ or numerical analyses $^8$ in the framework of  the
initial value  formalism.

 The general spherically symmetric  line element can be written as

$$ds^2= -\alpha^2dt^2+adr^2+br^2[d\theta^2+\sin^2 (\theta )d\phi^2],\eqno(1)$$
where $0<\phi  <2\pi $ and $0<\theta <\pi $ are the standard angle variables.

The initial data are prescribed by giving the spatial geometry at t = 0,
i.e., by specifying the functions $a$ and $b$ and by giving the extrinsic
curvature (which we assume to be traceless, $K_i^i=0$)

$$K_r^r=-2K_{\theta }^{\theta }= -2K_{\phi  }^{\phi  }=-{2\partial_tR
\over \alpha R}, \eqno(2)$$
where the areal radius $R$ is defined as

$$R=r\sqrt{b} .\eqno(3)$$
A useful concept is that of the  mean curvature of a centered sphere
in a Cauchy slice,

$$p = {2\partial_rR \over \sqrt{a} R}.\eqno(4)$$

It is important to note that each of these three quantities, $K_r^r$, $R$ and
$p$, are geometric three-scalars and as such do not depend on the coordinates
chosen on the three-slice.

The spherically symmetric Einstein equations  can be written as
2 initial constraints, one evolution equation and the lapse equation. We will
only consider the constraints in this article.

The  Hamiltonian constraint  can be written as

$$\partial_r p = -\sqrt{a} \Biggl\lbrace 8\pi \rho_0 + {3\over 4} (K_r^r)^2
+{3\over 4}p^2 -{1 \over R^2}\Biggr\rbrace ,\eqno(5)$$

while the momentum constraint reads

$$\partial_r K_r^r = -{3\over 2} p\sqrt{a} K_r^r -8\pi j_r.\eqno(6)$$

If  $T_{\mu }^{\nu }$ is the  energy-momentum tensor  of the matter field which
generates the spherical solution, then

$$\rho_0 = -T_0^0,~~~~j_r = -\alpha T^0_r\eqno(7)$$.

The   constraints (5, 6) possess a ``conserved'' quantity;

$$E = {p^2R^3\over 8} -{R\over 2} -  {R^3(K_r^r)^2\over 8 }
-2\pi\int_r^{\infty }\sqrt{a}\rho_0 R^3pd\tilde r - {1\over  4}
\int_r^{\infty }K_r^r \partial_{\tilde r}(K_r^rR^3)d\tilde r. \eqno(8)$$

$E$ is $r-$independent, $\partial_r E=0$. This can be proved directly by
differentiating  (8) and using the constraints (5, 6).

Assume that the initial data are asymptotically flat. Assuming a suitable
integrability of the current and
matter densities,
asymptotically the mean curvature $p$ approaches 0 as ${2 -4m/r\over r}$
while  the areal radius $R$ behaves like $r(1+{2m\over r})$, where $m$
is the ADM mass of the system. One then gets

$$E=-m.\eqno(9)$$

After a little algebra one can rewrite (8) as follows

$$m-\Bigl( \sqrt{{S\over 16\pi }} \Bigr) =
-{R^3\over 8}\theta (S) \theta'(S)
+2\pi\int_r^{\infty }\sqrt{a}\rho_0 R^3pd\tilde r + {1\over  4}
\int_r^{\infty }K_r^r \partial_{\tilde r}(K_r^rR^3)d\tilde r, \eqno(10)$$
where $S= 4\pi R^2$, $\theta (S)$ is the divergence of future-directed light
rays  outgoing from a sphere $S$ ($\theta = {1\over {\alpha }}{d\over
dt}\mid_{out} \ln S=p-K_r^r$) and  $\theta ' (S)$ is the divergence of
past-directed light rays   outgoing from $S$ (the convergence of
future-directed light rays ingoing from $S$) (  $\theta '= {-1\over {\alpha
}}{d\over dt}\mid_{in}\ln S =p + K_r^r$). Future trapped surfaces are those on
which $\theta(S)$ is negative, past trapped surfaces are those on which $\theta
'$ is negative.

The Penrose inequality is the statement that the left hand side of (10)
is positive on any apparent horizon, where future horizons are defined by
$\theta(S)$ vanishing and past horizons by $\theta '(S)$ being zero. Equation
(10), therefore,
would prove this inequality for spherically symmetric black holes
provided that the combination of the  two  integrals on the right hand side is
nonnegative.  We show that this is so if we impose very weak conditions on
the initial data.

Let us notice that, using the momentum constraint (6), the last integrand
can be written as

$$-8\pi R^3K_r^r j_r  . \eqno(11)$$

The two integrands appearing in (10) can now be written as

$$2\pi \int_r^{\infty }\sqrt{a}R^3\Bigl(
\rho_0  p  -  K_r^r {j_r\over \sqrt{a}} \Bigr) d\tilde r. \eqno(12)$$

and equation (10) becomes

$$m -  \sqrt{{S\over 16\pi }}   =
-{R^3\over 8}\theta (S) \theta'(S) +
2\pi \int_r^{\infty }\sqrt{a}R^3\Bigl(
\rho_0  p  -  K_r^r {j_r\over \sqrt{a}} \Bigr) d\tilde r. \eqno(13)$$

We   assume the weak energy  condition$^{10}$, which implies that

$$ \rho_0 \ge  {\mid j_r \mid\over \sqrt{a}} \eqno(14)$$
holds {\it outside} of the trapped surfaces in the data. We do not place any
restriction on the matter in the interior.

 This gives us
that the integrand in (13) is bounded from below by

$$ \sqrt{a}R^3 \inf [\rho_0 ( p  -  K_r^r),~~ \rho_0 (p  +  K_r^r)]=
\sqrt{a}R^3\inf [\rho_0 \theta ,~~  \rho_0  \theta '].$$

Consider the outermost future trapped surface, the (future) apparent horizon,
call it $S$. Let us assume that $S$ is outside the outermost past trapped
surface. In other words, we assume $\theta(S) = 0$ and that both $\theta$ and
$\theta '$ are positive outside $S$. From (14) we also have that $\rho_0 \ge 0$
in the same region.  Then (13) implies that

$$m-  \sqrt{{S\over 16\pi }}   \ge  2\pi \int_r^{\infty }R^3d\tilde r\sqrt{a}
\inf [\rho_0 \theta ,~~\rho_0  \theta ']\ge 0
\eqno(15)$$
Of course, an identical argument works if the outermost trapped surface is a
past apparent horizon.Thus we have demonstrated that the Penrose  inequality
holds for the outermost horizon. Let us stress that the only assumption we make
is that the matter satisfies the weak energy condition in the {\it exterior}
region. This may be important if one wishes to consider quantum effects which
may lead to violations of the energy conditions. These effects will tend to be
significant only in the strong field regions, i.e., inside the horizon.

An immediate consequence of (15) is that if $S$ is  the outermost apparent
horizon then (assuming the energy condition) the ADM  mass $m$ is greater than
(or , in vacuum, equal to) $R$/2. Hence (15) constitutes a necessary condition
for the  formation of a  trapped surface. It complements previously obtained
results; we failed in   $^1$ to find a necessary condition for non-symmetric in
time initial data$^{11}$.

We can also derive a sufficient condition using this same analysis.
In fact, one can show (assuming, as always, the weak energy condition)
 that outside a region with
trapped surfaces

$$R\theta  \le 2, ~~~~R\theta ' \le 2.$$

This allows one to get an estimate, using the same technique as before, that

$$2\pi \int_r^{\infty }\sqrt{a}R^3\Bigl(
\rho_0  p  -  K_r^r {j^r\over \sqrt{a}} \Bigr) d\tilde r
\le M_e $$

where $M_e(S)$ is the external rest mass $\int_{out ~S}\rho_0 dV  $
outside a surface $S$.  Thus, a sufficient condition is:

\vskip 0.4cm
{\it If $m\ge {R(S)\over 2}+M_e(S)$ then S is a trapped surface.}
\vskip 0.4cm
Unfortunately, the analysis of moment of time symmetry data
suggests that the above condition may, perhaps, never be satisfied.

At moment of time symmetry, both $j^r$ and $K_r^r$ are identically zero and
equation (13) reduces to
$$m -  \sqrt{{S\over 16\pi }}   =
-{R^3\over 8}p^2 +
2\pi \int_r^{\infty }\sqrt{a}R^3
\rho_0  p  d\tilde r. \eqno(16)$$
As before, we can show that the volume integral is less than $M_e$.

 Hence, in the case of symmetric in time data, we always have that $m\le
{R\over 2}+M_e$, independent of the existence or not of trapped surfaces. We do
not know whether this condition can be violated in the case of non-symmetric in
time data.

Gibbons$^3$ proposed a modified form of (15) for charged matter,

$$m-\Bigl( \sqrt{{S\over 16\pi }}+q^2\sqrt{{\pi \over S}}\Bigr) \ge 0,
  \eqno(17)$$
where $q$ is a global charge.
(17) can be proved assuming that all charged matter is enclosed within
the outermost trapped surface. In this case the energy density can be written
as
$\rho =\rho_e + \rho_m $, that is, it splits into a purely  electrostatic
(monopole)  part,

$$\rho_e= {q^2\over 18\pi R^4 }\eqno(18) $$

and the  remaining matter density,  represented by $\rho_m $, which  we assume
satisfies the energy condition (14). The integral of the electrostatic part
can be performed explicitly, to give

$$\int_r^{\infty }  {q^2\partial_rR \over 2R^2} d\tilde r
=q^2\sqrt{{\pi \over S}}.\eqno(19)$$

This gives the equation

$$m-\Bigl( \sqrt{{S\over 16\pi }}+q^2\sqrt{{\pi \over S}}\Bigr) =
-{R^3\over 8}\theta (S) \theta'(S)
+2\pi \int_r^{\infty }\sqrt{a}\rho_m R^3pd\tilde r + {1\over  4}
\int_r^{\infty }K_r^r \partial_{\tilde r}(K_r^rR^3)d\tilde r. \eqno(20)$$

The rest of the reasoning is exactly the same as before, so that  finally
one obtains (17).

We would like to point out that   the Penrose inequality ((15) with $q=0$) is
always   true, irrespective of whether the matter is charged or not and
independent of the detailed distribution of charged  matter.
The modification proposed by Gibbons for charged systems, however,  requires a
closer examination. We assumed above that the charged matter is  contained
entirely inside the outermost trapped surface. Can the above result  be true
without this condition?  Equation  (20) strongly suggests a negative  answer.
It is clear that eqn.(20) is valid even when the charge extends outside the
horizon if we understand $\rho_m$ to represent $\rho_m = \rho - \rho_e$
where $\rho_e$ is defined by (18).
In this case, however, $\rho_e$ is an overestimate of the electrostatic energy
and $\rho_m$ underestimates the matter energy. Even if the matter were to
satisfy the weak energy condition we have no reason to assume that the
unphysical $\rho_m$ would do so. It is easy to imagine an initial
geometry with an apparent horizon, where  a large  charge is carried by  matter
outside the region with  trapped surfaces. It should be simple to arrange
things so that the integrals  (20) are negative
thus violating the inequality (17).

On the other hand,  (17) will be true in the later stages of gravitational
collapse. One can show that the outermost  trapped surface
must move outward faster than light rays, thus swallowing more and more
of the matter; ultimately  the right hand side of (20) becomes negligible
and therefore the  left hand side of   (20) should equal zero.
  This in turn  suggests  that the areal radius
of trapped surfaces should grow (since both global charge $q$ and mass $m$
are conserved) during an evolution. That is  indeed true; the directional
derivative  of the area of the outermost trapped surface $S$ is equal to

$$ R^2K_r^r\sqrt{a}(V-{\alpha \over \sqrt{a}}),$$

where $V$ is   velocity of $S$  and ${\alpha \over \sqrt{a}}$ the velocity
of outgoing light rays. At the future horizon $K_r^r $ is strictly
positive: $K_r^r = {1\over 2}(-\theta +\theta ')= \theta ' /2$.

Let us comment on our assumptions concerning the focusing properties of the
space-time geometry. The condition that $\theta '$ is strictly positive
outside a sphere of vanishing $\theta (S)$ means that white holes (if there are
any ) are hidden inside $S$.  Ingoing light rays  are always convergent (i. e.,
$\theta '>0$ )  in any  realistic gravitational collapse that develops
from smooth initial data. Or, in other words; if at a some time $t$
the geometry of a collapsing system contains a surface with vanishing
$\theta '$, then the past history of that system must contain a singularity.
In the case of spherical  symmetry   the existence of a singularity
follows directly from the the Raychaudhuri equations$^{10}$, while
in the general nonspherical case one can invoke the Penrose - Hawking
singularity theorems. And conversely, one can prove that if smooth initial
data with strictly positive $\theta '$ give rise to a smooth evolution  then
ingoing light rays are always convergent, $\theta '>0$ on all future Cauchy
slices. It is reasonable, therefore,
to assume that $\theta '$ is always positive.

Acknowledgement. This work is partially supported by the grant of Komitet
Bada\'n Naukowych 2526/92.
\vfill \eject
\centerline{\bf References}

$^1$ P. Bizo\'n, E. Malec and N. \'O Murchadha: Phys. Rev. Lett., {\bf 61,}
1147
(1988); Classical and Quantum Gravity, {\bf 6}, 961 (1989); Classical
and Quantum  Gravity, {\bf 7}, 1953 (1990).

$^2$ R. Penrose: N Y Acad Sci {\bf 224}, 125 (1973).

$^3$ G. Gibbons  in {\it Global Riemannian Geometry} ed. N J Willmore and
N J Hitchin, Cambridge University Press (1984).

$^4$ R. Penrose:  Rivista Nuovo Cimento {\bf 1}, 251 (1969).

$^5$ M. Ludvigsen and J. A. Vickers: J Phys {\bf A16}, 3349 (1983).

$^6$ R. Bartnik: Journal of  Differential Geometry {\bf 37}, 31(1993).

$^7$ J. Jezierski: Class Quantum Gravity {\bf 6}, 1535 (1988).

$^8$ J. Karkowski, E. Malec and Z. \'Swierczy\'nski: Class Quantum Gravity
{\bf 10} 1361 (1993).

$^9$ J. Karkowski,  P. Koc and Z. \'Swierczy\'nski 1993  {\it Penrose
inequality
for gravitational waves}

$^{10}$ S. W. Hawking and G. F. R.  Ellis    {\it The large scale structure of
space-time}, Cambridge University Press (1973).

$^{11}$ Zannias has derived a necessary condition with moving matter
in the the special case that the matter is all moving in the same direction in
 T. Zannias, Phys. Rev. {\bf D 45}, 2998 (1992).

 \end